\documentclass[a4paper,12pt]{article}
\usepackage{amsmath,amsfonts,amssymb,cite}
\usepackage{graphicx}

\begin{document}

\begin{center}
{\Large\bf Holographic Estimates of the Deconfinement Temperature}
\end{center}

\begin{center}
{\large S. S. Afonin, A. D. Katanaeva
}
\end{center}

\begin{center}
{\small V. A. Fock Department of Theoretical Physics,
Saint-Petersburg State University, 1 ul. Ulyanovskaya, 198504,
Saint Petersburg, Russia}
\end{center}

\begin{abstract}
The problem of self-consistent estimates of the deconfinement
temperature $T_c$ in the framework of the bottom--up holographic
approach to QCD is scrutinized. It is shown that the standard soft
wall model gives $T_c$ for the planar gluodynamics around 260~MeV
in a good agreement with the lattice data. The extensions of the
soft wall model adjusted for descriptions of realistic meson
spectra result in a broad range of the predictions. This
uncertainty is related with a poor experimental information on the
radially excited mesons.
\end{abstract}

\section{Introduction}

The ongoing experiments on heavy ion collisions at ALICE (the
Large Hadron Collider at CERN), RHIC (the Brookhaven National
Laboratory) and planned experiments at FAIR (GSI) have caused an
increasing interest in the theoretical study of the QCD phase
diagram. One of the primary questions is to calculate the critical
temperature $T_c$ at which hadronic matter is supposed to undergo
a transition to a deconfined phase~\cite{Smilga:1996cm}. It is
believed that this transition played a crucial role in forming our
visible universe in the first few microseconds of its existence.
Under the real conditions of the present heavy ion collisions and
in the early universe, the influence of the finite baryon density
is negligible and can be set to zero in a first approximation.
This case is accessible for lattice simulations with an almost
realistic quark mass spectrum. Recently such lattice calculations
of $T_c$ have reached unprecedented levels of accuracy (see, e.g.,
the discussions in Ref.~\cite{Borsanyi:2010bp}).

From the theoretical side, one of the central problems in studying
the QCD matter under extreme conditions consists in derivation of
a relation between the deconfinement temperature and known hadron
parameters. Some time ago, a rather simple and elegant method for
calculating $T_c$ was proposed by Herzog~\cite{Herzog} within the
bottom--up approach to QCD. Based on the insight of
Ref.~\cite{Witten:1998zw} as regards confinement in
$\mathcal{N}=4$ super Yang--Mills theory on a sphere, the
deconfinement was related to a Hawking--Page phase transition
between a low temperature thermal AdS space and a high temperature
black hole in the AdS/QCD models. This interpretation proved to be
fully consistent with all large-$N_c$ field theory expectations.
The application of this idea to the hard~\cite{hw} and soft
wall~\cite{sw} models of AdS/QCD resulted in a semi-quantitative
prediction of $T_c$ as a function of the $\rho$-meson mass
$m_{\rho}$.

The phenomenological fits and comparison with the lattice data
performed in Ref.~\cite{Herzog} are rather short and disputable.
The agreement of obtained $T_c$ with a lattice result seems to be
a coincidence as we will show. In view of many new lattice data
and recent developments in the AdS/QCD models, we find it useful
to reconsider and extend Herzog's analysis. This will be the main
goal of our work. First, it will be argued that $m_{\rho}$ seems
not to be a good quantity for predicting $T_c$ in the holographic
models. One should use the parameters describing the whole tower
of radially excited states. Second, the dependence of $T_c$ on the
choice of experimental data and on hypotheses about missing data
will be analyzed. This discussion has a generic character. Third,
we will demonstrate that for the descriptions of realistic spectra
one should extend the soft wall model of Ref.~\cite{sw}. The
analysis of~\cite{Herzog} will be applied for a couple of such
extensions. At the end we discuss some other problems related with
the holographic calculations of deconfinement temperature.

\section{Hawking--Page phase transition}

We briefly recall the essence of Herzog's analysis~\cite{Herzog}.
Under some set of assumptions, the gravitational part of the
action of the dual theory takes the form
\begin{equation}
\label{1}
I=\kappa\int d^4xdze^{-\Phi}\sqrt{g},
\end{equation}
where the dilaton profile $\Phi=0$ for the hard wall
(HW)~\cite{hw} and $\Phi=az^2$ for the soft wall (SW)~\cite{sw}
model. The gravitational part~\eqref{1} yields the leading
contribution to the full action in the large-$N_c$ counting
($\kappa\sim N_c^2$ while the mesonic part scales as $N_c$). The
part~\eqref{1} is the same for AdS with a line element
\begin{equation}
\label{2}
ds^2=\frac{L^2}{z^2}\left(dt^2-d\vec{x}^2-dz^2\right),
\end{equation}
and for AdS with a black hole with the line element
\begin{equation}
\label{3}
ds^2=\frac{L^2}{z^2}\left(f(z)dt^2-d\vec{x}^2-\frac{dz^2}{f(z)}\right),
\end{equation}
where $f(z)=1-(z/z_h)^4$ and $L$ denotes the AdS radius. The
Hawking temperature is related to the black hole horizon $z_h$ via
the relation $T=1/(\pi z_h)$.

The free action density $V$ in the field theory is identified with
the regularized action $I$. The regularization consists in
dividing out by the volume of $\vec{x}$ space and imposing an
ultraviolet cutoff $z=\epsilon$. For thermal AdS, the energy
density reads
\begin{equation}
\label{4}
V_{\text{Th}}(\epsilon)=\kappa L^5\int_0^\beta dt\int_\epsilon^{z_0}e^{-\Phi}z^{-5}dz,
\end{equation}
while for the case of a black hole in AdS, the density becomes
\begin{equation}
\label{5}
V_{\text{BH}}(\epsilon)=\kappa L^5\int_0^{\pi z_h}dt\int_\epsilon^{\min(z_0,z_h)}e^{-\Phi}z^{-5}dz.
\end{equation}
The infrared cutoff $z_0$ is finite in the HW model~\cite{hw} and
$z_0=\infty$ in the SW one~\cite{sw}. The two geometries are
compared at a radius $z=\epsilon$ where the periodicity in the
time direction is locally the same, i.e. $\beta=\pi
z_h\sqrt{f(\epsilon)}$. The order parameter for the phase
transition is defined by the difference
\begin{equation}
\label{6}
\Delta V = \lim_{\epsilon\rightarrow\infty}\left(V_{\text{BH}}(\epsilon)-V_{\text{Th}}(\epsilon)\right).
\end{equation}
The thermal AdS is stable when $\triangle V>0$, otherwise the
black hole is stable. The Hawking--Page phase transition occurs at
a point where $\triangle V=0$. The corresponding critical
temperature of the HW model is
\begin{equation}
\label{7}
T_c=\frac{2^{1/4}}{\pi z_0}.
\end{equation}
For the SW model one arrives at
\begin{equation}
\label{8}
\Delta V = \frac{\pi\kappa L^5}{2z_h^3}\left[e^{-y_h}(y_h-1)+\frac12-y_h^2\int\limits_{y_h}^{\infty}\frac{dt}{t}e^{-t}\right],
\end{equation}
where $y_h\equiv az_h^2$. Numerical calculation
gives
\begin{equation}
\label{9}
T_c\approx0.49\sqrt{a}.
\end{equation}

The prediction for the deconfinement temperature was made
in~\cite{Herzog} from matching to the experimental $\rho$-meson
mass $m_\rho=776$~MeV~\cite{pdg}. The vector spectrum of HW
model is defined by roots of Bessel function $J_0(m_n z_0)=0$. The
first zero of $J_0$ yields $m_\rho\approx2.405/z_0$, hence
$z_0\approx(323~\text{MeV})^{-1}$. Thus the prediction is
\begin{equation}
\label{10}
T_c\approx0.157m_\rho=122~\text{MeV}.
\end{equation}
The vector spectrum of the SW model has a linear Regge-like
form~\cite{sw}
\begin{equation}
\label{11}
m_n^2=4a(n+1), \qquad n=0,1,2,\dots.
\end{equation}
Identifying the ground ($n=0$) state with the $\rho$-meson, one
obtains $\sqrt{a}=338$~MeV and
\begin{equation}
\label{12}
T_c\approx0.246m_\rho=191~\text{MeV}.
\end{equation}
The value~\eqref{12} lies very close to one of the lattice
predictions~\cite{karsch}. Based on this observation, it was
concluded that improved description of the spectrum in the SW
model (compared with the HW one) seems to entail the improved
prediction for $T_c$~\cite{Herzog}.

\section{Predictions: problems and Uncertainties}

The first uncertainty comes from the fact that one could consider
other types of particles in the AdS/QCD models (scalars,
axial-vectors etc.) which would lead to different predictions. One
can argue of course that the vector case looks the most
trustworthy in the holographic approach since the problem with
anomalous dimension of interpolating operator is absent due to
conservation of vector current. Confining ourselves to the sector
of light non-strange vector mesons, the second uncertainty arises
from the use of experimental value for $m_\rho$ in the SW model.
The spectrum of well-established and not confirmed $\rho$ and
$\omega$ mesons is shown in Table~1 and displayed graphically in
Figs.~1 and~2 respectively. It is well seen that the ground state
lies substantially lower than it is predicted by the averaged
linear trajectory. The identification of the slope with $m_\rho^2$
[as follows from~\eqref{11}] is therefore a crude approximation.
\begin{table}[ht]
\caption{\small The masses of known $\rho$ and $\omega$
mesons~\cite{pdg}. Not well-established but observed by several
groups resonances are marked by asterisk. The question mark stays
at the states observed by single group or states poorly
established (section "Further States" in Particle
Data~\cite{pdg}).} \vspace{-0.1cm}
\begin{center}
\begin{tabular}{|c|c||c|c|}
\hline
Name & Mass & Name & Mass\\
\hline
$\rho(770)$ & $776$ & $\omega(782)$ & $783$ \\
$\rho(1450)$ & $1465\pm 25$ & $\omega(1420)$ & $1400-1450$ \\
$\rho(1570)$ *& $1570\pm 36$ & $\omega(1650)$ & $1670\pm30$ \\
$\rho(1700)$ & $1720\pm20$ & $\omega(1960)$ ?& $1960\pm25$ \\
$\rho(1900)$ *& $1909\pm17$ & $\omega(2205)$ ?& $2205\pm30$ \\
$\rho(2000)$ ?& $2000\pm30$ & $\omega(2290)$ ?& $2290\pm20$ \\
$\rho(2150)$ *& $2155\pm21$ & $\omega(2330)$ ?& $2330\pm30$ \\
$\rho(2270)$ ?& $2265\pm40$ & & \\
\hline
\end{tabular}
\end{center}
\end{table}

\begin{figure}
\center{\includegraphics[scale=0.4]{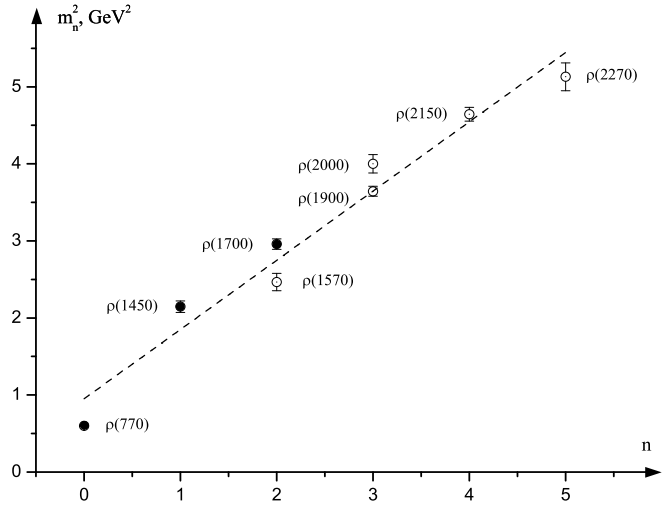}} \vspace{-1cm}
\caption{\small Our assignment of radial number $n$ to the
$\rho$-mesons from Table~1. The well-established states are
filled.}
\end{figure}
\begin{figure}
\center{\includegraphics[scale=0.4]{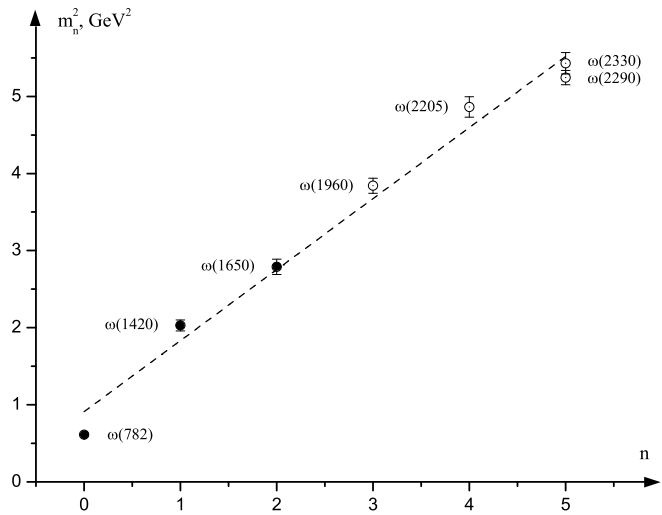}}
\vspace{-1cm}
\caption{\small Our assignment of $n$ to the $\omega$-mesons
from Table~1.}
\end{figure}

A more reasonable strategy for making estimates consists in the
direct use of the slope which is controlled by the parameter $a$
in~\eqref{11}. But the accurate extraction of $a$ from the data in
Table~1 is not so straightforward as it may seem. First of all,
more than a half of states are not well confirmed or poorly known.
The unconfirmed states have a different degree of belief. For
instance, a concrete mean mass for $\rho(1900)$ is even not given
in the Particle Data~\cite{pdg} (we have used an experimental
result of Ref.~\cite{Aubert:2007ym}). The question arises whether
we should use these states for fitting the linear trajectory and
if we should, then which weight must be ascribed to each of these
states in averaging procedure. The use of three well-established
states for drawing the linear trajectory is also questionable.
First of all, the ground vector states lie noticeably below the
linear trajectory. This situation is common for the vector
quarkonia~\cite{AP}. Since we do not know reliably the underlying
reason, it could make sense to exclude the ground states from the
trajectory. The first radially excited states --- $\rho(1450)$ and
$\omega(1420)$ --- are situated unnaturally higher the linear
trajectory and, in fact, have a peculiar status. The matter is
that they represent just names for broad resonance regions rather
than well defined resonances~\cite{pdg}. As is emphasized in
Particle Data, the mass $m_{\rho(1450)}=1465\pm25$~MeV is "only an
educative guess". This resonance seems to have some admixture of
the strange quark (enlarging its mass) and its decays show
characteristics of hybrids~\cite{pdg}. The situation with
$\omega(1420)$ is similar. The resonances $\rho(1700)$ and
$\omega(1650)$ also have much less clear status than $\rho(770)$
and $\omega(782)$. In addition, they are often interpreted as the
$D$-wave vector states. In the compilations~\cite{ani,bugg}, they
give rise to the second radial vector trajectory (the first one
contains the $S$-wave states). Such an interpretation is typical
for semi-relativistic potential models~\cite{godfrey}. According
to this physical picture, the given states do not represent the
radial excitations of $\rho$ and $\omega$. Thus, we see that all
well-established vector mesons have specific problems which do not
allow one to make a reliable fit.

In spite of all these uncertainties, if we look at tentative
linear trajectories for various light non-strange
mesons~\cite{ani,bugg}, a remarkable feature emerges: The slope is
approximately universal quantity, i.e. it weakly depends on
quantum numbers of trajectory. This observation is a strong
argument in favor of the hypothesis (inspired by the hadron string
models~\cite{nambu}) that the slope is mainly determined by the
gluodynamics. On the other hand, within the SW holographic
model~\cite{sw}, the slope is also universal for mesons of any
spin and parities and even for glueball trajectories. The use of
the universal slope for estimates of $T_c$ partly resolves the
problem of dependence of the predicted value for $T_c$ on the
quantum numbers of mesons under consideration. According to the
review~\cite{bugg}, the mean slope of radial trajectories is (in
terms of~\eqref{11}): $4a=1.14$~GeV$^2$. Substituting this value
to~\eqref{9} we obtain
\begin{equation}
\label{13}
T_c\approx263~\text{MeV}.
\end{equation}
This estimate gives much larger value for $T_c$ than predicted
by~\eqref{12}. The discrepancy is caused by the fact that the
relation~\eqref{11} yields a much heavier "$\rho$-meson", about
1068~MeV for the real phenomenological slope.

As follows from our discussions, if one normalizes not to the
physical $\rho$-meson mass but to the best fit, the predicted
$T_c$ is increased. A similar situation takes place in the HW
model. The best global fit is achieved at the cutoff value
$z_0=(346~\text{MeV})^{-1}$~\cite{hw}. This would correspond to a
heavier $\rho$-meson, $m_\rho=832$~MeV~\cite{hw}, and a higher (in
comparison with~\eqref{10}) value for the deconfinement
temperature, $T_c=131$~MeV.

Comparison of theoretical predictions for the deconfinement
temperature with the corresponding lattice results deserves a
special consideration. The lattice simulations measure $T_c$ in
units of some dimensional quantity. The standard choice for this
quantity is the string tension $\sigma$ which is obtained from the
linear behavior of the potential between two static quarks at a
large separation, $U(r)=\sigma r$ at large distance $r$. The
standard value of $\sigma$ used in the most of lattice simulations
is $\sqrt{\sigma}=420$~MeV. The prediction~\eqref{12} practically
coincides with the lattice result of Ref.~\cite{karsch}. This
coincidence was the main quantitative result of Herzog's
analysis~\cite{Herzog}. However, a closer look at the related
paper~\cite{cheng} shows that the obtained lattice result is
$T_c/\sqrt{\sigma}=0.419(6)$. After that a larger value for
$\sigma$ was used, $\sqrt{\sigma}=460$~MeV, for predicting $T_c$.
With the standard value of $\sigma$, the result of
Ref.~\cite{cheng} (and of~\cite{karsch}) is $T_c\approx175$~MeV.

In the presence of massive quarks, the deconfinement phase
transition represents a crossover occurring in some range of
temperatures. The exact position of this crossover depends on the
observable used to define it (this is a general feature of all
crossover transitions). Some time ago, the value of $T_c$ on the
lattice with physical quarks was vastly debated in the literature.
Some measurements gave the range $180-200$~MeV
(e.g.~\cite{cheng,Bazavov:2009zn}), another measurements resulted
in $150-170$~MeV
(e.g.~\cite{Aoki:2006br,Bazavov:2011nk,Bernard:2004je}), see
Ref.~\cite{Borsanyi:2010bp} for a detailed discussion. But after
the recent progress in extrapolating to the continuum limit and to
the physical light quark masses, different lattice methods have
converged to the range $150-170$~MeV~\cite{Borsanyi:2010bp}.

This interval for $T_c$, however, does not suit a comparison with
the estimates following from Herzog's analysis. We wish to clearly
emphasize this point. According to the philosophy of AdS/QCD
correspondence, the gravitational part of the holographic
action~\eqref{1} is dual to pure gluodynamics in the large-$N_c$
limit. Hence, the predicted value for $T_c$ must be compared with
the lattice results for gluodynamics (i.e. with non-dynamical
quarks) extrapolated to large $N_c$. Such an extrapolation was
carried out in Ref.~\cite{Lucini:2012wq}. The result is:
$T_c/\sqrt{\sigma} = 0.5949(17)+0.458(18)/N_c^2$. With
$\sqrt{\sigma}=420$~MeV, this extrapolation leads to $T_c=250$~MeV
in the large-$N_c$ limit. For $N_c=3$, one has $T_c=271$~MeV. This
interpolation agrees with the lattice simulations for $SU(3)$
Yang--Mills theory in Refs.~\cite{Boyd:1996bx}
($T_c/\sqrt{\sigma}=0.629(3)$, $T_c=264(1)$~MeV)
and~\cite{Iwasaki:1996ca} ($T_c/\sqrt{\sigma}\approx0.65$,
$T_c\approx273$~MeV).

Finally we see that the prediction~\eqref{13} of the SW model
looks much more successful than the prediction~\eqref{12} claimed
in the original paper~\cite{Herzog}. In addition, the
self-consistency of the method is improved: The deconfinement
phase transition is of the 1-st order in the $N_c\geq3$
gluodynamics and its strength grows with
$N_c$~\cite{Lucini:2013qja}. This means that in the limit
$N_c\rightarrow\infty$, the transition becomes of the same type as
the Hawking--Page phase transition.

\section{Deconfinement temperature in modified SW Models}
\subsection{The generalized SW  model}

The linear vector spectrum~\eqref{11} of the standard SW
model~\cite{sw} contains a strictly fixed intercept. If we
interpolate the points in Fig.~1 or~2 by the linear function, the
realistic spectrum will differ from the pattern~\eqref{11}. Let us
generalize the spectrum~\eqref{11},
\begin{equation}
\label{14}
m_n^2=4a(n+1+b), \qquad n=0,1,2,\dots,
\end{equation}
where the parameter $b$ will control the intercept for
phenomenological spectra. The generalization of the SW
model~\cite{sw} which leads to the vector spectrum~\eqref{14} is
known~\cite{genSW}. It requires the following form for the dilaton
profile in~\eqref{1},
\begin{equation}
\label{15}
\Phi=az^2-2\ln{U(b,0;az^2)},
\end{equation}
here $U$ denotes the Tricomi hypergeometric function ($U(0,0;x)=1$). The
deconfinement temperature will depend now not only on the slope
parameter $a$ but also on the intercept parameter $b$. Below we
briefly study this dependence.

The expression~\eqref{8} is generalized to
\begin{multline}
\label{16}
\Delta V=\frac12\pi\kappa L^5 y_h\left[\frac{U^2(b,0;0)}{4y_h^2}-\int\limits_{y_h}^{\infty}\frac{dt}{t^3}U^2(b,0;t)e^{-t}\right]=\\
\frac{\pi\kappa L^5}{2z_h^3}\left\{e^{-y_h}U^2(b,0;y_h)(y_h-1)+2 b y_he^{-y_h}U(b,0;y_h)U(1+b,1;y_h)+\right.\\
\frac{1}{2\Gamma(1+b)}-y_h^2\int\limits_{y_h}^{\infty}\frac{dt}{t}e^{-t}\left[U^2(b,0;t)+4b U(1+b,1;t)U(b,0;t)+\right.\\
\left.\left.2b^2U^2(1+b,1;t)+2b(1+b)U(2+b,2;t)U(b,0;t)\right]\right\}.
\end{multline}
The equation $\Delta V=0$ yields $y_h$ at a given $b$, after that
$T_c$ is determined from the relation
$T_c=(\pi\sqrt{y_h/a})^{-1}$. The dependence of $T_c/\sqrt{a}$ on
$b$ in the interval $-0.5\leq b\leq 0.5$ is displayed in Fig.~3. For
$b\gtrsim-0.3$, this dependence is practically linear,
\begin{equation}
\label{17}
T/\sqrt{a}=0.496+0.670b.
\end{equation}
\begin{figure}
\vspace{-2cm}
\center{\includegraphics[scale=1]{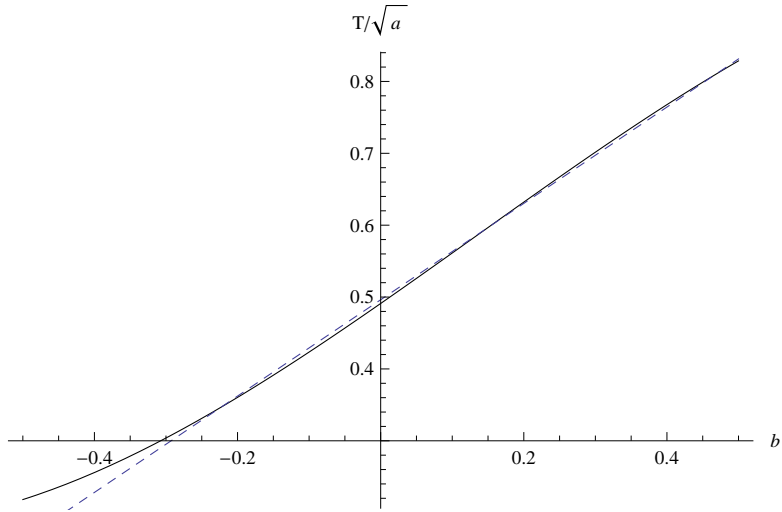}}
\caption{\small The dependence of $T_c/\sqrt{a}$ on $b$ (see text). The dotted line shows the interpolation~\eqref{17}.}
\end{figure}

One can fix some value of $T_c$ and find a parametric curve on the
$(a,b)$ plane corresponding to the given $T_c$. For the
value~\eqref{13}, this curve is shown in Fig.~4. The points on (or
close to) this curve correspond to the choices of $a$ and $b$ at
which the generalized SW model reproduces more or less the
physical value of $T_c$ in gluodynamics. It looks really
surprising that the simplest version of the SW model ($b=0$)
introduced in Ref.~\cite{sw} belongs to the physically acceptable
region on the $(a,b)$ plane.
\begin{figure}[t]
\center{\includegraphics[scale=1]{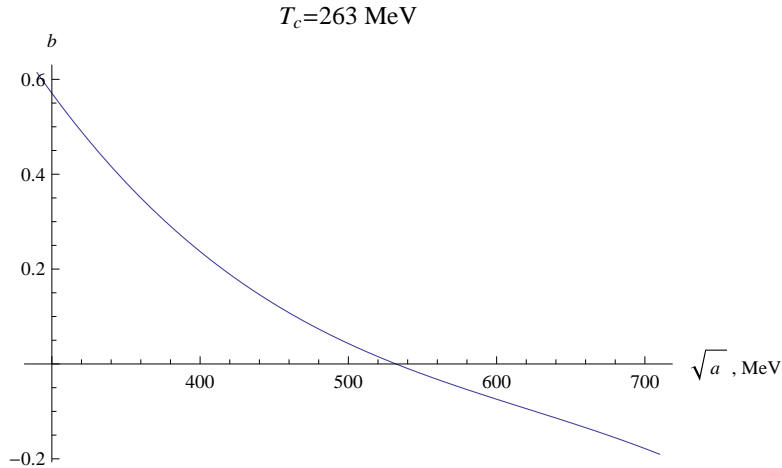}}
\caption{\small The parametric curve on the $(a,b)$ plane
corresponding to $T_c=263$~MeV.}
\end{figure}

Consider, in the spirit of Ref.~\cite{Herzog}, the prediction of
$T_c$ from a realistic vector spectrum. For this purpose, we need
to extract the parameters $a$ and $b$ from the $\rho$ or $\omega$
spectrum in Table~1. As we discussed in Sect.~3, the extracted
values will strongly depend on the choice of data and on the
weight of each state in the fit. In this situation, the account
for the experimental errors in the mass determination is not very
informative since, in practice, such errors are subleading in the
final fit. We will take the central values of the masses and the
predicted $T_c$ should be regarded as an estimate. We analyze how
different hypotheses on the choice of data for interpolating the
linear trajectory influence on the predicted value of $T_c$. The
results are summarized in Table~2.

Some comments are in order. We considered three hypotheses. In the
first one, only the well-established states are used. In this
case, the predicted value of $T_c$ lies a bit below the interval
$150-170$~MeV given by lattices with dynamical light quarks and
the $\rho$ and $\omega$ sectors yield close results. Next we add
the poorly known states except the following resonances: the $n=5$
excitations as the least established states, the $\rho(1570)$ and
$\rho(1900)$ since they represent $\phi\pi$ states with a large
hidden strange component appearing jointly in a certain fit of
experimental data~\cite{Aubert:2007ym}. Here the $\rho$ and
$\omega$ sectors result in quite different predictions. Most
likely, this is related with an insufficient accuracy of the
experimental data. A definite choice for the $\rho$ and $\omega$
states leads to a prediction of $T_c$ in the interval
$250-270$~MeV expected in gluodynamics. This choice constitutes
our third hypothesis. Such a possibility is interesting because
the requirement of a correct reproduction of $T_c$ could serve as
a guide for the prediction (confirmation) of new resonances
within the SW model.
\begin{table}[ht]
\caption{\small Some predictions for $T_c$ basing on Table~1,
Figs.~1 and~2, relations~\eqref{14} and~\eqref{16} (see text).}
\vspace{-0.1cm}
\begin{center}
\begin{tabular}{|c|c|c|c|}
\hline
Particle & Radial states & $m_n^2$, GeV$^2$ & $T_c$, MeV\\
\hline
$\rho$ & $n=0,1,2$ & $1.18(n+0.61)$ & $143$ \\
$\omega$ & $n=0,1,2$ & $1.09(n+0.66)$ & $149$ \\
$\rho$ & $n=0,1,2,3,4$ & $0.99(n+0.89)$ & $207$ \\
$\omega$ & $n=0,1,2,3,4$ & $1.03(n+0.74)$ & $166$ \\
$\rho$ & $n=0,1,2,4,5$ & $0.88(n+1.12)$ & $270$ \\
$\omega$ & $n=1,2,3,4$ & $0.95(n+1.04)$ & $255$ \\
\hline
\end{tabular}
\end{center}
\end{table}

\subsection{The SW model with the UV cutoff}

The linear radial Regge trajectory is only an approximation to the
observable spectrum. The attempts to introduce non-linearities
into the SW model lead usually to models admitting only numerical
treatment. We will consider the model of Ref.~\cite{cut} which can
be solved analytically. The non-linearity of the SW spectrum is
introduced in~\cite{cut} via imposing the ultraviolet (UV) cutoff.
An heuristic physical motivation is rather simple. In the UV
regime, QCD represents a weakly coupled gauge theory, hence,
according to the ideas of holographic duality, its probable
holographic dual should be in the strong coupling regime. This
makes questionable the applicability of a semiclassical
approximation to the dual theory when $z\rightarrow0$. The
introduction of UV cutoff is a crude way for avoiding this
problem. The vector spectrum becomes non-linear, it is given by
zeros of the Tricomi function $U(-m_n^2/(4a),0;aL^2)$~\cite{cut},
where $L$ is the AdS radius and the cutoff is imposed (without
loss of generality) at $z_{UV}=L$. The spectrum has the form
$m_n^2=4af_n(aL^2)$ with $f_n$ representing a function of the
cutoff value. For example,
$f_n(1)=\{1.57,2.84,4.05,5.22,6.37,\dots\}$.

The comparison of the model with the real spectra was not
performed in Ref.~\cite{cut}. For our purposes, we partly analyze
the ensuing phenomenology. It is convenient to rewrite the
spectrum~\eqref{11} in units of the ground mass,
\begin{equation}
\label{18}
m_n^2=m_0^2\{1,2,3,\dots\}.
\end{equation}
In these notations, the $\rho$ and $\omega$ spectra from Table~1
are
\begin{equation}
\label{19}
m_{\rho,n}^2=m_\rho^2\{1,3.6,4.9,?,7.7,8.5\},\quad m_{\omega,n}^2=m_\omega^2\{1,3.3,4.5,6.3,7.9,?\}.
\end{equation}
The non-zero cutoff does not allow to improve the agreement
of~\eqref{18} with the experimental patterns~\eqref{19}. However,
consider the axial-vector $a_1$-mesons. The Particle
Data~\cite{pdg} cites one well-established resonance of the mass
$1230\pm40$~MeV and three poorly known states with the masses
$1647\pm22$, $1930^{+30}_{-70}$, and $2270^{+55}_{-40}$~MeV. The
spectrum can be written as
\begin{equation}
\label{20}
m_{a_1,n}^2=m_{a_1}^2\{1,1.8,2.5,3.4\}.
\end{equation}
The model prediction in the example above is
\begin{equation}
\label{21}
aL^2=1:\qquad m_n^2=m_0^2\{1,1.8,2.6,3.3,4.1,\dots\}.
\end{equation}
We see that the spectra~\eqref{20} and~\eqref{21} are very close,
i.e. the SW model with the UV cutoff is able to provide an
accurate description for the axial-vector spectrum. This supports
the arguments of Ref.~\cite{cut} that the UV cutoff mimics the
chiral symmetry breaking.

We will use this property of the model under consideration to
estimate $T_c$ from the axial-vector sector. The extension
of~\eqref{8} to the case of finite UV cutoff $z=\epsilon$ is
\begin{equation}
\label{22}
\Delta V = \frac12\pi\kappa L^5z_h\left[\int\limits_{y_\epsilon}^{y_h}\frac{dt}{t^3}e^{-t}-
\sqrt{1-\frac{y_\epsilon^2}{y_h^2}}\int\limits_{y_\epsilon}^{\infty}\frac{dt}{t^3}e^{-t}\right],
\end{equation}
where $y_h\equiv az_h^2$, $y_\epsilon\equiv a\epsilon^2$. The
equation $\Delta V = 0$ allows one to find $y_h$ from a fixed
value of $y_\epsilon$. Taking the fit considered above,
$y_\epsilon=1$, we obtain numerically $y_h\approx1.40$ that for
the mean value of the radial slope, $4a=1.14$~GeV$^2$, corresponds
to $T_c\approx144$~MeV. This prediction practically coincides with
that of the $\rho$-meson sector in Table~2.

\section{Discussions}

The contribution of the chiral symmetry breaking to the full
holographic action scales as $N_c$. Since the prediction of $T_c$
comes from the gluonic part~\eqref{1} scaling as $N_c^2$, one
could naively think that the axial-vector spectrum is equally good
for predicting $T_c$ and estimate a discrepancy with the vector
case at the $1/N_c$ level. This expectation is of course not
correct. Even the rough analysis of Ref.~\cite{Herzog} would give
a unrealistically large difference
$T_c^{(a_1)}/T_c^{(\rho)}=m_{a_1}^2/m_\rho^2\approx2.5$. For a
more consistent prediction we should extract the parameters $a$
and $b$ from the linear fit of the $a_1$ trajectory and find $T_c$
from the generalized SW model. The result is
$a\approx0.30$~GeV$^2$, $b\approx0.25$ which leads to
$T_c\approx363$~MeV. The large enhancement of predicted $T_c$
occurs due to a large value of $b$
--- this is clear from Fig.~3 and the approximate
relation~\eqref{17}. The prediction for the deconfinement
temperature from the axial-vector sector is surprisingly close to
the estimates from the vector one if the SW model with the UV
cutoff is exploited.

Since the results of lattice simulations are usually given in
units of the string tension $\sigma$, the possible errors in
determination of $\sigma$ entail some uncertainty in the lattice
predictions for $T_c$. This source of uncertainty could be avoided
if theoretical predictions were also expressed in terms of
$T_c/\sqrt{\sigma}$. Unfortunately, such an expression is model
dependent. For instance, if we assume the string (flux tube)
picture of mesons, assume that the meson string is of the
Nambu--Goto type and identify the tension of relativistic string
with the tension of non-relativistic linear potential, then the
slope~\eqref{11} is given by~\cite{Zwiebach,baker}
\begin{equation}
\label{23}
4a=2\pi\sigma,
\end{equation}
i.e. $\sqrt{a}=\sqrt{\pi/2}\sqrt{\sigma}\approx1.25\sqrt{\sigma}$
in~\eqref{9} and in similar formulas. In particular, the
result~\eqref{9} becomes $T_c/\sqrt{\sigma}\approx0.62$ which
agrees well with the lattice predictions (see Sect.~3). The
phenomenological mean slope~\cite{bugg} 1.14~GeV$^2$ yields
$\sqrt{\sigma}\approx426$~MeV if we use~\eqref{23}. This value is
also close to many lattice measurements,
$\sqrt{\sigma}\approx420$~MeV. The assumptions above lead thus to
a reasonable picture. We give an heuristic derivation of the
slope~\eqref{23} in the appendix.

The requirement of the existence of a non-zero deconfinement
temperature restricts the possible form for the dilaton background
$\Phi$ in the holographic action~\eqref{1}. It is easy to check
that if the sign of $\Phi$ is changed then $\Delta V<0$
in~\eqref{8} at all temperatures, i.e. the model is always in the
deconfined phase. This conclusion seems to contradict to the
Sonnenschein criterion of confinement~~\cite{Sonnenschein:2000qm}
based on the Wilson loop area law for the confinement of strings.
According to this criterion, the time--time metric component
$g_{00}$ should satisfy the conditions
\begin{equation}
\label{24}
\partial_z(g_{00})|_{z=z_0}=0,\qquad g_{00}|_{z=z_0}\neq0.
\end{equation}
The AdS metric~\eqref{2} does not satisfy~\eqref{24}. The dilaton
profile $e^{-\Phi}$, however, can be rewritten as a part of the
metric which becomes asymptotically ($z\rightarrow0$) AdS. The
choice $\Phi=az^2$ results in monotonically decreasing $g_{00}$,
the condition~\eqref{24} cannot be fulfilled, while the choice
$\Phi=-az^2$ provides a non-trivial minimum for $g_{00}$ matching
the confinement criterion~\eqref{24}. This property was exploited
in Ref.~\cite{Andreev:2006ct} for a derivation of the linear
confinement potential from the holographic approach and later
triggered an active use of the SW models with inverse dilaton
profile (see, e.g.,~\cite{Zuo:2009dz,schmidt}) in spite of a
formal existence of massless vector mode~\cite{Karch:2010eg}. Thus
we see that the black hole and Wilson loop criteria for
confinement are in conflict in the simplest version of the SW
model. A resolution of this puzzle would be interesting. An
obvious possibility consists in a modification of the dilaton
profile $e^{-az^2}$ with preserving its infrared asymptotics.

In the gravitational action~\eqref{1}, the form of the dilaton
background is the same as in the SW model of Ref.~\cite{sw}. It
should be noted that in reality this represents a rather strong
assumption as long as the SW model has not been derived from any
string theory. One can simply imagine a situation when this
assumption is violated. Indeed, suppose that the planar
gluodynamics is dual to the closed string sector of a full dual
theory. Then constructing an effective gravity dual in AdS$_d$
space one commonly arrives at the expression
\begin{equation}
\label{25}
I=\int d^dx e^{-2\Phi}\sqrt{g}L_{\text{grav}}+\int d^dx
e^{-\Phi}\sqrt{g}L_{\text{matter}},
\end{equation}
in which the condensate $\Phi$ of a massless scalar field (called
the dilaton) controls the string coupling. In realistic models of
physics, the gravitational part of~\eqref{25} may contain a
dilaton potential possessing some minimum, say at $\Phi=az^2$.
Comparing~\eqref{25} with~\eqref{1} we should then conclude that
the dilaton contribution is rescaled by the factor of 2 in the
gluonic part in comparison with the mesonic part of the action.
This means that the slope parameter should be rescaled as
$a\rightarrow2a$ in making predictions for $T_c$. The
prediction~\eqref{9} becomes
$T_c\approx0.49\sqrt{2a}\approx0.70\sqrt{a}$ yielding
$T_c\approx372$~MeV instead of~\eqref{9}. Note that if we use the
fit of the original analysis~\cite{Herzog}, where $m_\rho^2$ is
identified with the slope, the prediction~\eqref{12} is
$T_c\approx0.348m_\rho=270$~MeV, which lies amusingly close to the
lattice predictions in the $SU(3)$ Yang--Mills
theory~\cite{Boyd:1996bx,Iwasaki:1996ca}. This agreement hints at
the idea that a SW-like model leading to the vector spectrum
$m_n^2=4a(n+1/2)$ would be successful in predicting $T_c$ on the
base of~\eqref{25}. Using some modifications of the holographic
prescriptions, such a variant of the SW model was proposed in
Ref.~\cite{schmidt}. Its spectrum reads\footnote{In essence, this
is the spectrum of Ademollo--Veneziano--Weinberg dual
amplitude~\cite{avw}.} $m^2=4a(n+(L+J)/2)$, where $J$ is the total
spin and $L$ denotes the orbital momentum of a quark--antiquark
pair. Here the vector spectrum ($L=0$, $J=1$) is degenerate with
the scalar one ($L=1$, $J=0$) and is automatically shifted with
respect to the axial-vector spectrum ($L=1$, $J=1$).

In the case of free intercept, the relation~\eqref{25} suggests to
rescale the dilaton~\eqref{15},
\begin{equation}
\label{26}
\Phi=2az^2-4\ln{U(b,0;az^2)},
\end{equation}
that renders~\eqref{16} into
\begin{equation}
\label{27}
\Delta V=\frac12\pi\kappa L^5 y_h\left[\frac{U^4(b,0;0)}{4y_h^2}-\int\limits_{y_h}^{\infty}\frac{dt}{t^3}U^4(b,0;y)e^{-2y}\right].
\end{equation}
The condition $\Delta V=0$ gives now another predictions for
$T_c$. For the input data in Table~2, these predictions are shown
in Table~3.
\begin{table}[htb]
\caption{\small The predictions for the deconfinement temperature
based on inputs from Table~2 in the case of replacement~\eqref{16}
by~\eqref{27}.} \vspace{-0.1cm}
\begin{center}
\begin{tabular}{|c|c|c|c|}
\hline
Particle & Radial states & $m_n^2$, GeV$^2$ & $T_c$, MeV\\
\hline
$\rho$ & $n=0,1,2$ & $1.18(n+0.61)$ & $159$ \\
$\omega$ & $n=0,1,2$ & $1.09(n+0.66)$ & $170$ \\
$\rho$ & $n=0,1,2,3,4$ & $0.99(n+0.89)$ & $299$ \\
$\omega$ & $n=0,1,2,3,4$ & $1.03(n+0.74)$ & $199$ \\
$\rho$ & $n=0,1,2,4,5$ & $0.88(n+1.12)$ & $400$ \\
$\omega$ & $n=1,2,3,4$ & $0.95(n+1.04)$ & $381$ \\
\hline
\end{tabular}
\end{center}
\end{table}

Finally we see that predictions for the deconfinement temperature
depend strongly on assumptions as regards the possible origin of
the SW model from a dual string theory.

\section{Conclusions}

We have analyzed in detail various aspects of the prediction for
the deconfinement temperature $T_c$ from the bottom--up
holographic models of QCD. It was argued that the predicted $T_c$
must refer to deconfinement phase transition in the pure
gluodynamics. The agreement of the prediction of the simplest soft
wall model~\cite{sw} with the recent lattice results looks
impressive. We have also shown that if the soft wall model is
accommodated for a description of realistic vector spectra, the
predicted $T_c$ becomes ambiguous because of lack of sufficient
amount of reliable experimental data on the radially excited light
mesons. The use of well-established states results in $T_c$ close
to the crossover transition in the lattice simulations with
dynamical quarks.

The arising relations between parameters of the observed radial
trajectories of light mesons and the deconfinement temperature in
the planar QCD represent a curious theoretical result of the
holographic approach. The fact that in many cases these relations
agree well with the lattice results means that the holographic
trick seems to pass an important phenomenological test. On the
other hand, the requirement of a reasonable prediction for $T_c$
can serve as a strong restriction on the possible variants of the
holographic models. These restrictions may be useful for
predicting new resonances.

\section*{Acknowledgments}

The author acknowledges Saint-Petersburg State University for a research grant
11.38.189.2014. The work was also partially supported by the
RFBR grant 13-02-00127-a.

\section*{Appendix}

There exists a simple heuristic way for the derivation of the
linear radial trajectory from the semiclassical flux tube model
for the light mesons~\cite{Shifman:2005zn}. It leads to the wrong
slope; nevertheless, it is occasionally used in the literature and
in discussions. We briefly reproduce this derivation and then
correct it.

Consider a thin gluon flux tube of length $r$ stretched between
massless quark and antiquark. The energy of the system (the meson
mass) is
$$
M=2\sqrt{p}+\sigma r,
$$
where $p$ denotes the momentum of quarks oscillating in the linear
confinement potential. The tension of the flux tube (the string
tension) is defined by $\sigma=M/l$, here $l$ means the maximal
quark separation. Impose the Bohr--Sommerfeld quantization
condition on the quark momentum,
$$
\int\limits_0^l pdr=\pi(n+b), \qquad n=0,1,2,\dots.
$$
The constant $b$ depends on the boundary conditions ($b=1/2$ for
the centrosymmetrical potentials). A trivial integration results
in the relation
$$
M^2=4\pi\sigma(n+b).
$$
The slope obtained is twice the slope of the Nambu--Goto
string~\eqref{23}. The source of the discrepancy lies in the
unphysical assumption as regards the massless quarks which, was
used in the essentially non-relativistic derivation.

Let us introduce the quark masses $m_1$ and $m_2$ and consider the
system in the rest frame of the quark 1. The energy of the system
is
$$
M=m_1+\sqrt{p^2+m_2^2}+\sigma r.
$$
Assume that $m_2$ is much less than the typical momentum $p$,
$m_2^2/p^2\ll 1$. Repeating the derivation above, we get
$$
(M-m_1)^2\simeq2\pi\sigma(n+b).
$$
Insofar as $m_1,m_2\ll M$ in the light mesons, we can safely
neglect the small mass contributions stemming from $m_1$ and
$m_2$. The final Regge-like formula has the correct slope.

\end{document}